\documentclass[prd,10pt,aps,nofootinbib,twocolumn,floatfix,amssymb,amsmath,amsfonts,longbibliography]{revtex4-2}

\usepackage{epsfig}
\usepackage{slashed}

\newcommand{\beqn}{\begin{eqnarray}}
\newcommand{\eeqn}{\end{eqnarray}}
\newcommand{\beqs}{\begin{subequations}}
\newcommand{\eeqs}{\end{subequations}\\[-2mm]\noindent}
\newcommand{\eq}[1]{(\ref{#1})}

\newcommand{\cS}{{S}}

\renewcommand{\P}{{\cal{P}}}

\newcommand{\NH}{{\mathrm{H}}}
\newcommand{\T}{{\cal{T}}}

\newcommand{\PT}{{\mathcal{PT}}}

\newcommand{\bs}{\boldsymbol}

\def\bbbone{{\mathchoice {\rm 1\mskip-4mu l} {\rm 1\mskip-4mu l} {\rm 1\mskip-4.5mu l} {\rm 1\mskip-5mu l}}}

\usepackage{xcolor}

\allowdisplaybreaks
\begin{document}

\title{IR/UV mixing from local similarity maps of scalar non-Hermitian field theories}

\author{Maxim N.~Chernodub}
\affiliation{Institut Denis Poisson UMR 7013, Universit\'e de Tours, 37200 France}
\email{maxim.chernodub@idpoisson.fr}

\author{Peter Millington}
\affiliation{Department of Physics and Astronomy, University of Manchester, Manchester M13 9PL, United Kingdom\\
School of Physics and Astronomy, University of Nottingham, University Park, Nottingham NG7 2RD, United Kingdom}
\email{peter.millington@manchester.ac.uk}

\begin{abstract}
We propose to ``gauge'' the group of similarity transformations that acts on a space of non-Hermitian scalar theories. We introduce the ``similarity gauge field'', which acts as a gauge connection on the space of non-Hermitian theories characterized by (and equivalent to a Hermitian) real-valued mass spectrum. This extension leads to new effects:~if the mass matrix is not the same in distant regions of space, but its eigenvalues coincide pairwise in both regions, the particle masses stay constant in the whole spacetime, making the model indistinguishable from a standard, low-energy and scalar Hermitian one. However, contrary to the Hermitian case, the high-energy scalar particles become unstable at a particular wavelength determined by the strength of the emergent similarity gauge field. This instability corresponds to momentum-dependent exceptional points, whose locations cannot be identified from an analysis of the eigenvalues of the coordinate-dependent squared mass matrix in isolation, as one might naively have expected. For a doublet of scalar particles with masses of the order of 1 MeV and a similarity gauge rotation of order unity at distances of 1 meter, the corrections to the masses are about $10^{-7}\,\mathrm{eV}$, which makes no experimentally detectable imprint on the low-energy spectrum. However, the instability occurs at $10^{18} \, \mathrm{eV}$, suggestively in the energy range of detectable ultra-high-energy cosmic rays, thereby making this truly non-Hermitian effect and its generalizations of phenomenological interest for high-energy particle physics.
\\
~
\\
\footnotesize{This is an author-prepared post-print of \href{https://doi.org/10.1103/PhysRevD.105.076020}{Phys.\ Rev.\ D {\bf 105} (2022) 076020}, published by the American Physical Society under the terms of the \href{https://creativecommons.org/licenses/by/4.0/}{CC BY 4.0} license (funded by SCOAP\textsuperscript{3}).}
\end{abstract}

\date{\today}

\maketitle


\section{Introduction}

The quantum mechanics of systems with non-Hermitian Hamiltonians has attracted significant attention, since the realization that Hermiticity can be superseded by other antilinear symmetries~\cite{Mannheim:2015hto}, while still guaranteeing real energy eigenvalues and a unitary evolution~\cite{Bender:2002vv}.  In many examples, the antilinear symmetry at work is the combined action of parity $\mathcal{P}$ and time-reversal $\mathcal{T}$~\cite{Bender:1998ke} (for an introduction, see Ref.~\cite{Bender:2005}).  More generally, such models fall into the class of pseudo-Hermitian quantum theories~\cite{Mostafazadeh:2001jk, Mostafazadeh:2001nr, Mostafazadeh:2002id}.

For many pseudo-Hermitian quantum theories, there exists a non-unitary, similarity transformation that maps the non-Hermitian Hamiltonian to an isospectral Hermitian one~\cite{Mostafazadeh:2002id}.  In the case of quantum mechanics, this transformation can be time dependent (for recent studies, see Refs.~\cite{Fring:2020mhu, fring2021infinite}).  In the case of quantum field theories, this similarity transformation can be made local.  The latter is the focus of this work, wherein we describe a local similarity transformation of an archetypal non-Hermitian scalar field theory.  This allows us to introduce an associated vector field, which we refer to as a ``similarity gauge field'', which allows spatial variations of the parameters of the field theory to be connected.

Spacetime inhomogeneity of fundamental particle parameters (masses, couplings) can lead to various phenomena that differentiate one spacetime region from another~\cite{experimentreview}.\footnote{Here, we stress the word ``spacetime'', since we envisage not only space-dependent but also time-dependent phenomena.} For example, inhomogeneous masses can serve as external potentials acting on particle (field) constituents of the model. At the same time, varying couplings can enhance certain processes in one region of space and inhibit them in another region~\cite{Adams:2019kby}.

The parameters of the Standard Model of particle physics are known to be spatially uniform to a high degree in a vast local region of our universe~\cite{Barrow:1999qk}. Moreover, the time independence of various couplings, especially of the fine-structure constant, has been shown to hold to a high accuracy within billions of years~\cite{fine2020}. Therefore, any noticeable spacetime variation of the parameters of the Standard Model seems to be excluded. Our article challenges this by showing that weakly inhomogeneous mass parameters in a non-Hermitian theory can lead to a tiny effect on locally measured physical masses at low energies---thus being compatible with local mass measurements---while noticeably affecting the propagation of particles at very high energies. This behavior sounds puzzling and nontrivial, because it asserts that the high-energy, ultraviolet (UV) particle spectrum is affected by very slow, infrared (IR) spacetime variations of the mass matrix of the model. We will show that this merging of two different 
scales is a particular feature of a non-Hermitian model, originating in the viability of anti-Hermitian interactions. At the same time, its Hermitian analog---which possesses a similar spectrum and the same particle content---does not exhibit this exotic mixing of UV and IR scales.

This novel connection between weak spacetime variations of physical parameters and the UV dynamics could have interesting applications in the context of cosmology. For instance, it may provide a new mechanism by which IR physics is screened from (non-Hermitian) modifications to the gravitational dynamics, which instead affect only the UV physics (cf.~other screening mechanisms in, e.g., Refs.~\cite{Joyce:2014kja, Koyama:2015vza, Burrage:2017qrf}). Alternatively, in the presence of spacetime-dependent defects, such as domain walls or cosmic strings, and once extended to higher-spin fields, this effect could significantly alter the high-energy phenomenology, for instance, of cosmic rays from superconducting strings (see, e.g., Refs.~\cite{Hill:1986mn,Berezinsky:2009xf, Brandenberger:2019lfm}).

The remainder of this article is organized as follows. In Sec.~\ref{sec:model}, we begin by introducing the prototypical scalar non-Hermitian model on which our analysis will focus.  We describe the global similarity transformation that maps it to a corresponding Hermitian theory in Sec.~\ref{sec:global}, before generalizing this to the case of a local similarity transformation in Secs.~\ref{sec:local} and~\ref{sec:localmass}.  Doing so necessitates the introduction of a new vector field---the ``similarity gauge field''---and its impact upon the classical equations of motion and dispersion relations is analysed in Sec.~\ref{sec:eoms}. We include a comparison to the {\it ab initio} Hermitian case in Sec.~\ref{sec:Hermitiancomparison}. Before providing our conclusions in Sec.~\ref{sec:conc}, we further elucidate the impact of the similarity gauge theory by means of a physical example in Sec.~\ref{sec:physical_realization}.  Details of the operator-level formulation of the local similarity transformation are provided in the Appendix for completeness.


\section{Non-Hermitian PT-symmetric bosons}
\label{eq:global:bosons}


\subsection{Lagrangian}
\label{sec:model}

Following Ref.~\cite{Alexandre:2017foi}, we consider a minimal non-Hermitian scalar model, which contains a complex doublet
\beqn
\Phi = \begin{pmatrix}
\phi_1 \\ \phi_2
\end{pmatrix}\,,
\eeqn
whose squared mass matrix is non-Hermitian, having the form\footnote{This model was introduced in Ref.~\cite{Alexandre:2017foi} in the context of Noether's theorem, and similar self-interacting extensions of this model have been studied in the contexts of the Goldstone theorem and the Englert-Brout-Higgs mechanism~\cite{Alexandre:2018uol, Mannheim:2018dur, Alexandre:2018xyy, Alexandre:2019jdb, Fring:2019hue, Fring:2019xgw, Fring:2020bvr}.}
\beqn
M^2 = 
\begin{pmatrix}
\phantom{-} m_1^2 & m_5^2 \\[1mm]
- m_5^2 & m_2^2
\end{pmatrix}
\label{eq:mass:NH}
\eeqn
and describing mixing between the two complex scalar fields $\phi_1$ and $\phi_2$.  The squared mass matrix is skew symmetric, and there exists a matrix
\begin{equation}
    P=P^{-1}=\begin{pmatrix} 1 & 0 \\ 0 & -1\end{pmatrix}\,,
\end{equation}
which fulfills the pseudo-Hermiticity condition
\begin{equation}
    PM^2P=\big[M^2\big]^{\mathsf{T}}\,,
\end{equation}
where $\mathsf{T}$ indicates the matrix transpose. We should therefore expect that the squared mass matrix, although non-Hermitian, has a real eigenspectrum within certain regions of the parameter space. The squared mass eigenvalues are
\beqn
M_{\pm}^2 = \frac{1}{2} \left( m_1^2 + m_2^2 \pm \sqrt{\left( m_1^2 - m_2^2\right)^2 - 4 m_5^4} \right)\,.
\label{eq:M:pm}
\eeqn
These are real, and the model does not possess an instability in the region defined by the following constraints:
\beqs
\beqn
m_1^2 + m_2^2 & \geqslant & 0\,, 
\label{eq:stability:1}\\
m_1^2 m_2^2 + m_5^4 & \geqslant & 0\,,
\label{eq:stability:2}\\
\left( m_1^2 - m_2^2\right)^2 - 4 m_5^4 & \geqslant & 0\,. 
\label{eq:stability:3}
\eeqn
\label{eq:stability:NH} 
\eeqs
In Eq.~\eq{eq:stability:NH}, we have assumed that the squared masses $m_1^2$ and $m_2^2$ can take negative values. The conditions in Eq.~\eq{eq:stability:NH} then determine the region of the parameter space $(m_1^2,m_2^2,m_5^2)$ within which the $\PT$ symmetry is said to be unbroken~\cite{Alexandre:2017foi,Mannheim:2018dur,Fring:2019hue}. Hereafter, we will assume that the diagonal squared masses $m_1^2$ and $m_2^2$ are positive quantities, in which case the conditions~\eq{eq:stability:1} and \eq{eq:stability:2} are automatically satisfied, and we are left with only the condition~\eq{eq:stability:3}. In addition, without loss of generality, we will take $m_1^2>m_2^2$.

It is helpful to introduce a parameter~\cite{Alexandre:2017foi}
\beqn
    \eta\equiv \frac{2m_5^2}{m_1^2-m_2^2}\,,
\eeqn
which parametrizes the deviation from Hermiticity. At $\eta=0$, the two flavours decouple and the model is Hermitian. For $0<\eta<1$, the model is non-Hermitian but in the regime of unbroken $\mathcal{P}\mathcal{T}$ symmetry, where the squared mass eigenvalues are real. For $\eta>1$, the squared mass eigenvalues are complex, and the $\mathcal{P}\mathcal{T}$-symmetry is said to be broken. At $\eta=1$, the squared mass eigenvalues merge, we lose an eigenvector, and the squared mass matrix becomes defective. This {\it exceptional point}, a feature unique to non-Hermitian matrices, occurs at the boundary between the regimes of broken and unbroken $\mathcal{P}\mathcal{T}$ symmetry. A key finding of this article is that an analysis of the squared mass eigenvalues of a non-Hermitian, spacetime-dependent mass matrix is not sufficient to locate the exceptional points of the field theory. Instead, the spacetime dependence leads to momentum-dependent exceptional points, such that, for any given spacetime dependence, low-momentum modes remain in the regime of unbroken $\mathcal{PT}$ symmetry, whereas high-momentum modes are instead in the regime of broken $\mathcal{PT}$ symmetry.

The eigenvectors of the squared mass matrix~\cite{Alexandre:2017foi}
\beqs
\beqn
\mathbf{e}_+=N\begin{pmatrix} \eta \\ -1+\sqrt{1-\eta^2}\end{pmatrix}\,,\\
\mathbf{e}_-=N\begin{pmatrix} -1+\sqrt{1-\eta^2} \\ \eta \end{pmatrix}\,,
\eeqn
\eeqs
are not orthogonal with respect to the usual Dirac inner product. However, there exists an additional matrix $A$, given by~\cite{Alexandre:2020gah}
\beqn
A=\frac{1}{\sqrt{1-\eta^2}}\begin{pmatrix} 1 & \eta \\ -\eta & -1\end{pmatrix},\qquad A^2=\mathbb{I}\,,
\eeqn
with which the eigenvectors are orthogonal and with which we can define a positive-definite norm~\cite{Bender:2002vv}. We will refer to this as the $\mathcal{A}\mathcal{P}\mathcal{T}$ inner product:\footnote{In the case of $\mathcal{PT}$-symmetric quantum mechanics, this was introduced as the $\mathcal{C}\mathcal{P}\mathcal{T}$ inner product~\cite{Bender:2002vv}. However, $\mathcal{C}$ does not coincide with charge conjugation. To avoid confusion, this was denoted as the $\mathcal{C}'\mathcal{P}\mathcal{T}$ inner product in Ref.~\cite{Alexandre:2017foi}. Here, we use $\mathcal{A}\simeq \mathcal{C}'$ with $A=\left(C'\right)^{\mathsf{T}}$ from Ref.~\cite{Alexandre:2020gah}.}
\beqs
\label{eq:matrix_model_CpPT}
\beqn
\mathbf{e}_{\pm}^{\mathcal{A}\mathcal{P}\mathcal{T}}\cdot \mathbf{e}_{\pm}\equiv \mathbf{e}_{\pm}^*\cdot P\cdot A\cdot \mathbf{e}_{\pm}=1\,,\\
\mathbf{e}_{\pm}^{\mathcal{A}\mathcal{P}\mathcal{T}}\cdot \mathbf{e}_{\mp}\equiv \mathbf{e}_{\pm}^*\cdot P\cdot A\cdot \mathbf{e}_{\mp}=0\,,
\eeqn
\eeqs
fixing the normalization~\cite{Alexandre:2017foi}
\begin{equation}
    N=\left(2\eta^2-2+2\sqrt{1-\eta^2}\right)^{-1/2}\,.
\end{equation}
Notice that the $\mathcal{A}\mathcal{P}\mathcal{T}$ norm is ill defined as we approach the exceptional point $\eta\to 1$, since $|N|^2\to \infty$.

It will prove convenient in the analysis that follows to define a parameter
\beqn
\kappa\equiv \frac{1}{\sqrt{1-\eta^2}}=\frac{m_1^2 - m_2^2}{\sqrt{(m_1^2 - m_2^2)^2 - 4 m_5^4}}\,,
\eeqn
such that $\kappa=1$ corresponds to the Hermitian theory of two decoupled complex scalar fields and $\kappa>1$ corresponds to the unbroken $\mathcal{P}\mathcal{T}$-symmetric regime. The exceptional points lie at $\kappa\to \infty$, and $\kappa\in\mathbb{C}$ signals the regime of broken $\mathcal{P}\mathcal{T}$ symmetry.

The squared mass matrix is diagonalized by a similarity transformation $\mathcal{S}$ of the form
\beqn
M^2_{\rm diag}=\cS_{\rm H}^{-1}M^2 \cS_{\rm H}\,,
\eeqn
with
\beqn
\label{eq:S:eta}
\cS_{\rm H} &=&N\begin{pmatrix} \eta & -1+\sqrt{1-\eta^2} \\  -1+\sqrt{1-\eta^2} & \eta \end{pmatrix}\nonumber\\&=&
\frac{1}{\sqrt{2}}
\left( 
\begin{array}{cc}
\sqrt{\kappa + 1}     &  -\sqrt{\kappa - 1} \\[1mm]
-\sqrt{\kappa - 1}     & \sqrt{\kappa + 1}
\end{array}
\right)\,,
\label{eq:cS:xi}
\eeqn
giving $M^2_{\rm diag}={\rm diag}(M_+^2,M_-^2)$.\footnote{Relative to Ref.~\cite{Alexandre:2020gah}, we have introduced an overall minus sign into the definition of $S_{\rm H}$ by convention, so that the diagonal entries are positive in the $\mathcal{PT}$-symmetric regime.} The transformation is parametrized by the single real-valued, coordinate-independent parameter~$\kappa$ with $\cS_{\rm H}^\dagger \equiv \cS_{\rm H} $ and $\cS_{\rm H}^\dagger \neq \cS_{\rm H}^{-1}$. This similarity matrix is related to the $A$ matrix of the $\mathcal{A}\mathcal{P}\mathcal{T}$ transformation via
\beqn
\cS^2_{\rm H}=A P\,,
\eeqn
as is known for $\mathcal{PT}$-symmetric theories (see also the ``$V$-norm'' of Ref.~\cite{Mannheim:2017apd}).

It will prove convenient to write the similarity matrix in the form
\beqn
\cS_{\rm H}=e^{-\xi_{\rm H}\sigma_1}=\begin{pmatrix}
\phantom{-} \cosh \xi_{\rm H} & - \sinh \xi_{\rm H} \\
- \sinh \xi_{\rm H} & \phantom{-} \cosh \xi_{\rm H} \\
\end{pmatrix}\,,
\label{eq:S2:global}
\eeqn
where $\sigma_1$ is the first Pauli matrix and
\beqn
\xi_{\rm H}\equiv \frac{1}{2}{\rm arctanh}\, \eta=\frac{1}{2}{\rm arccosh}\, \kappa\,.
\label{eq:xiNHdef}
\eeqn
This makes the relation $\cS ^{-1}_{\rm H}(\xi_H) = \cS_{\rm H} (- \xi_H)$ apparent.

Having now outlined the non-Hermitian structure of the model, we can turn our attention to the Lagrangian of the scalar field theory.  We want to build our Lagrangian out of irreducible, scalar representations of the Lorentz group, as we do for Hermitian field theories. However, since the Hamiltonian is non-Hermitian, so too is the generator $P_0$ of the Poincar\'{e} group. We therefore have two copies of the algebra of the Lorentz group---one constructed from $P_0$ and the other from its Hermitian conjugate---and these algebras are disconnected, since $[P_0,P_0^{\dag}]\neq 0$. Hence, if we are to have canonical dynamics, we must build the Lagrangian out of scalar representations of only one of these Poincar\'{e} algebras, with both the field $\Phi$ and its conjugate momentum $\Pi$ transforming with respect to the same generator of time translations, $P_0$ say. Notice that $\Pi\neq \dot{\Phi}^{\dag}$, since $\Phi^{\dag}$ will evolve with respect to $P_0^{\dag}$.

Following Ref.~\cite{Alexandre:2020gah}, the Lagrangian of interest therefore takes the form
\beqn
    \label{eq:L:prime}
    \mathcal{L}=\partial_{\mu}\tilde{\Phi}^{\dag}\partial^{\mu}\Phi-\tilde{\Phi}^{\dag}M^2\Phi\,,
\eeqn
where $\tilde{\Phi}^{\dag}\neq \Phi^{\dag}$ will be defined below. Note that we could equivalently have chosen to work with the algebra of $P_0^{\dag}$, amounting to placing the tildes on $\Phi$ rather than $\Phi^{\dag}$, giving
\beqn
    \tilde{\mathcal{L}}=\partial_{\mu}\Phi^{\dag}\partial^{\mu}\tilde{\Phi}-\Phi^{\dag}M^2\tilde{\Phi}\,.
\label{eq:L:boson}
\eeqn
However, this is equivalent to the transformation $ m_5^2\to - m_5^2$, and the sign of $ m_5^2$ is irrelevant, since observables consistent with the $\mathcal{PT}$ symmetry depend only on $m_5^4$~\cite{Alexandre:2017foi}.  To see this, it is convenient to rewrite the Lagrangian~\eq{eq:L:boson} in terms of the individual scalar fields $\phi_1$ and $\phi_2$:
\beqn
 {\cal L} & = & \partial_\mu\tilde{\phi}_1^* \partial^\mu\phi_1 + \partial_\mu\tilde{\phi}_2^*\partial^\mu\phi_2 
 \nonumber \\
 & & - m_1^2\tilde{\phi}^*_1\phi_1-m_2^2\tilde{\phi}_2^*\phi_2 - m_5^2(\tilde{\phi}_1^*\phi_2 -\tilde{\phi}_2^*\phi_1)\,,
\eeqn
wherein we see that swapping the tildes in the final $m_5^2$-dependent term amounts to an overall change of sign.
 
Since $\tilde{\Phi}^{\dag}\neq \Phi^{\dag}$, the Euler-Lagrange equations obtained by varying with respect to $\tilde{\Phi}^{\dag}$ and $\Phi$, along with their Hermitian conjugates, are mutually consistent (cf.~Ref.~\cite{Alexandre:2017foi}).  Specifically, we have (with the d'Alembert operator $\Box \equiv \partial_\mu \partial^\mu = \partial_t^2 - {\bs \nabla}^2$):
\beqs
\beqn
    \Box \phi_1+m_1^2\phi_1+ m_5^2\phi_2&=&0\,,\\
    \Box \phi_2+m_2^2\phi_2- m_5^2\phi_1&=&0\,,\\
    \Box \tilde{\phi}_1+m_1^2\tilde{\phi}_1- m_5^2\tilde{\phi}_2&=&0\,,\\
    \Box \tilde{\phi}_2+m_2^2\tilde{\phi}_2+ m_5^2\tilde{\phi}_1&=&0\,,
\eeqn
\label{eq:classical:free}
\eeqs
along with their complex conjugates. Notice again that the untilded and tilded equations differ by $ m_5^2\to- m_5^2$. The untilded and tilded fields are related via parity (see Ref.~\cite{Alexandre:2020gah}), as is necessary since the Hamiltonian (and the Lagrangian) is not invariant under parity. For the $c$-number fields, we have the transformations
\begin{subequations}
\beqn
\P: & \quad & \Phi(t,{\bs x} ) \to \Phi'(t,-{\bs x}) = e^{i \alpha} \sigma_3 \tilde{\Phi}(t,{\bs x} ), \quad \\[2mm]
\T: & \quad & \Phi(t,{\bs x} ) \to \Phi'(-t,{\bs x}) = e^{i \beta} \Phi^*(t,{\bs x} )\,,
\label{eq:PT:transformations:NH}
\eeqn
\end{subequations}
with arbitrary parameters $\alpha=0,\pi$ and $\beta \in {\mathbb R}$, and where $\sigma_3$ is the third Pauli matrix. Notice that the scalar field $\phi_1$ transforms under the parity inversion $\P$ as a genuine scalar, whereas the scalar field $\phi_2$ behaves as a pseudoscalar. We can readily confirm that the Lagrangian \eqref{eq:L:prime} is $\mathcal{PT}$ symmetric.

There is an additional discrete symmetry---which we will call the $\mathcal{A}$ symmetry---which manifests at the level of the squared mass matrix as the invariance
\beqn
A M^2 A=M^2.
\eeqn
In terms of the fields themselves, this is effected as
\beqn
\mathcal{A}:\quad \left\{
\begin{array}{rcl}
\phi_i(t,{\bs x}) \, \to\, \phi_i'(t,{\bs x}) & = & A_{ij} \phi_j(t, {\bs x}), \quad \\[2mm]
\tilde{\phi}^{\dag}_i(t,{\bs x}) \to \tilde{\phi}_i^{\prime\dag}(t,{\bs x}) & = & A_{ji}\tilde{\phi}_j^{\dag}(t, {\bs x})\,.
\end{array}
\right.
\eeqn

In addition to the discrete spacetime symmetries described above, the model is also invariant under the global U(1) transformation
\beqn
\mbox{U(1)}: \quad \left\{
\begin{array}{rcl}
\Phi(t,{\bs x}) \to \Phi'(t,{\bs x}) = e^{i \gamma} \Phi(t, {\bs x}), \quad \\[2mm]
\tilde{\Phi}^{\dag}(t,{\bs x}) \to \tilde{\Phi}^{\prime\dag}(t,{\bs x}) = e^{-i \gamma} \tilde{\Phi}^{\dag}(t, {\bs x})\,,
\end{array}
\right.
\label{eq:U1}
\eeqn
which rotates the phases of both complex scalar fields $\phi_1$ and $\phi_2$ by the single real-valued phase factor $\gamma$.


\subsection{Global similarity transformation}
\label{sec:global}

There is a continuous set of $\PT$-invariant Lagrangians that are physically equivalent in the sense that they all possess the same physical spectrum. These models are related to each other by the similarity transformation $\mathcal{S}$, under which
\beqn
\Phi \to \cS  \Phi, \quad 
\tilde{\Phi}^\dagger \to \tilde{\Phi}^\dagger \cS ^{-1}\,,
\label{eq:tranfs:tilded}
\eeqn
where the similarity matrix $\cS$, parametrized by a single real-valued scalar parameter $\xi$, is given in Eq.~\eq{eq:S2:global}. The transformation of the tilded field~\eq{eq:tranfs:tilded} can be written in the following form:~$\tilde{\Phi} \to \cS ^{-1} \tilde{\Phi}$. (Recall that $S=S^{\dag}$ is Hermitian.)

One can show that the similarity transformation~\eq{eq:tranfs:tilded} with the matrix~\eq{eq:S2:global}, characterized by the specially fixed global parameter $\xi_{\rm H}$, as defined in Eq.~\eqref{eq:xiNHdef}, maps the original non-Hermitian model~\eq{eq:L:prime} to a Hermitian model of two non-interacting scalar fields with the constant masses~\eq{eq:M:pm} (see Ref.~\cite{Alexandre:2020gah}):
\beqn
\mathcal{L}_{\rm H}=\partial_{\mu}\Phi^{\dag}\partial^{\mu}\Phi-\Phi^{\dag}M^2_{\rm diag}\Phi\,,
\eeqn
wherein we have dropped the now redundant tildes on the conjugated fields.


\subsection{Local similarity transformation}
\label{sec:local}

Under a local transformation with $\xi \equiv \xi(x)$, the spacetime derivatives of the doublet scalar fields change as
\beqs
\beqn
\partial_\mu \Phi &&\to \partial_\mu \big( \cS  \Phi \big) \equiv \cS  \Big[\partial_\mu + \cS^{-1} \big(\partial_\mu \cS\big) \Big] \Phi\,,\\
\partial_\mu \tilde{\Phi}^{\dag} &&\to \partial_\mu \big( \tilde{\Phi}^{\dag}\cS^{-1} \big) \equiv  \tilde{\Phi}^{\dag}\Big[\overset{\leftarrow}{\partial}_\mu +  \big(\partial_\mu\cS^{-1}\big) \cS \Big]\cS ^{-1}\,. \qquad
\eeqn
\eeqs
In order to support the local similarity invariance, we introduce a new vector field $C_\mu$, which promotes the usual derivative to the covariant one, i.e., $\partial_\mu \to D_\mu$, with
\beqn
D_\mu = \partial_\mu - {\cal C}_\mu \equiv \bbbone \, \partial_\mu + \sigma_1 C_\mu 
\equiv 
\begin{pmatrix}
\partial_\mu & C_\mu \\
C_\mu & \partial_\mu 
\end{pmatrix}
\,.
\label{eq:D:mu}
\eeqn
Here, ${\cal C}_\mu \equiv - \sigma_1 C_\mu$ is the similarity vector ``gauge'' field, which evolves under the local gauge similarity transformation as follows:
\beqn
{\cal C}_\mu \to \cS  {\cal C}_\mu \cS^{-1} - \cS  \partial_\mu \cS^{-1}\,.
\label{eq:C2:trans}
\eeqn
However, we also need the tilde-conjugate covariant derivative
\beqn
\tilde{D}_\mu = \partial_\mu + {\cal C}_\mu \equiv \bbbone \, \partial_\mu - \sigma_1 C_\mu
\equiv 
\begin{pmatrix}
\partial_\mu & -C_\mu \\
-C_\mu & \partial_\mu 
\end{pmatrix}
\,,
\label{eq:D:tilde:mu}
\eeqn
and we see that the interactions of the similarity gauge field are non-Hermitian. We then have
\beqn
D_\mu \Phi \to \cS  D_\mu \Phi\,, 
\qquad
{\tilde D}_{\mu} {\tilde \Phi} \to \cS ^{-1} {\tilde D}_{\mu} \tilde{\Phi}\,,
\label{eq:DD:transformation}
\eeqn
and, consequently, $(\tilde{D}_\mu \tilde{\Phi})^{\dag} \to (\tilde{D}_\mu \tilde{\Phi})^{\dag} \cS^{-1}$. 
Here,  we have used the relations $\cS  \partial_\mu \cS^{-1} = - \cS^{-1} \partial_\mu \cS  = (\partial_\mu \cS^{-1}) \cS$ and $\cS  {\cal C}_\mu \cS^{-1} = \cS^{-1} {\cal C}_\mu \cS $.

This similarity transformation can also be expressed in terms of an operator $\hat{S}$ (see Ref.~\cite{Alexandre:2020gah}), and we include details of this in the case of the local transformation in the Appendix.

Using the explicit form of the similarity matrix~\eq{eq:S2:global}, as well as the transformation property~\eq{eq:C2:trans} of the matrix-valued similarity field ${\cal C}_\mu$, we obtain that the vector field $C_\mu$ transforms under the local similarity transformation~\eq{eq:C2:trans} as
\beqn
C_\mu \to C_\mu + \partial_\mu \xi\,.
\label{eq:C2:trans:inf}
\eeqn

Finally, we arrive at the gauge theory that is invariant under the local similarity transformation~\eq{eq:DD:transformation}:
\beqn
{\cal L}_C = 
[\tilde{D}_\mu \tilde{\Phi}]^\dagger D^\mu \Phi 
- \tilde{\Phi}^\dagger M^2 \Phi\,,
\label{eq:L:sim}
\eeqn
which can be written in the following explicit form:
\beqn
{\cal L}_C & = & (\partial_\mu \tilde{\phi}_1 - C_\mu \tilde{\phi}_2)^* (\partial^\mu \phi_1 + C^\mu \phi_2) \nonumber \\
& & + (\partial_\mu \tilde{\phi}_2 - C_\mu \tilde{\phi}_1)^* (\partial^\mu \phi_2 + C^\mu \phi_1) \nonumber \\
& & - m_1^2\tilde{\phi}^*_1\phi_1-m_2^2\tilde{\phi}^*_2\phi_2 - m_5^2(\tilde{\phi}_1^*\phi_2 - \tilde{\phi}_2^*\phi_1)\,.
\label{eq:L:sim:2}
\eeqn
Note that we do not treat the similarity gauge field as dynamical.

The similarity current is a non-Hermitian structure given by the variation of the non-Hermitian action with respect to the similarity gauge field:
\beqn
J^\mu_\NH & = & \frac{\delta S}{\delta C_\mu} = -\tilde{\Phi}^\dagger \sigma_1 D^\mu \Phi + [\tilde{D}^{\mu}\tilde{\Phi}]^{\dag}\sigma_1\Phi\nonumber \\
& = & -\tilde{\phi}_1^* \partial^\mu \phi_2 + \phi_1 \partial^\mu \tilde{\phi}^*_2
- \tilde{\phi}_2^* \partial^\mu \phi_1 + \phi_2 \partial^\mu \tilde{\phi}^*_1 \nonumber \\
& & - 2 C^\mu (\tilde{\phi}^*_1\phi_1 + \tilde{\phi}^*_2\phi_2)\,.
\label{eq:J:sim}
\eeqn

We remark that the global transformation with the coordinate-independent parameter~\eq{eq:xiNHdef} leaves the kinetic term in the Lagrangian~\eq{eq:L:sim} intact. This is because the non-diagonal part of both the usual~\eq{eq:D:mu} and tilded~\eq{eq:D:tilde:mu} covariant derivatives involve the matrix $\sigma_1$, which commutes with the similarity transformation, i.e., $\cS(\xi) \sigma_1 = \sigma_1 \cS(\xi)$.


\subsection{Local squared mass parameters}
\label{sec:localmass}

Now let us consider the non-Hermitian model with the coordinate-dependent mass term chosen in such a way that the eigenvalues of the mass matrix are kept the same at each point in spacetime. If we ignore the kinetic terms then the mass terms in the Hermitian counterpart of the non-Hermitian model would be given by the spacetime-independent quantities~\eq{eq:M:pm}. We show below that the presence of the kinetic terms makes the situation more complex:~the local nature of the transformation activates the non-unitary similarity transformation and leads to the appearance of the similarity gauge field~$C_\mu$. 

Indeed, consider the model with a vanishing similarity field $C_\mu = 0$ and globally constant squared mass eigenvalues. The latter requirement is correct provided the masses $m_1(x)$, $m_2(x)$, and $m_5(x)$ are spacetime dependent quantities subject to the condition that their combinations~\eq{eq:M:pm} are globally constant. This leads to two constraints on the three squared mass parameters:
\beqs
\label{eq:m0M0}
\beqn
m_1^2(x) + m_2^2(x) & = & M_0^2\,, \\{}
[m_1^2(x) - m_2^2(x)]^2 - 4 m_5^4(x) & = & m_0^4\,,
\eeqn
\eeqs
where $M_0$ and $m_0$ are the fixed parameters that enter the physical masses
\beqn
M_\pm^2 = \frac{M_0^2 \pm m_0^2}{2}\,.
\label{eq:Mpm:phys}
\eeqn

The most obvious way to proceed is to parametrize the mass matrix $M^2$ in terms of the single real function of spacetime coordinates
\beqn
\theta(x) = \arctan\frac{m_2(x)}{m_1(x)}\,.
\eeqn
The resulting squared mass matrix
\beqn
M^2(x) = \left( 
\begin{array}{cc}
M_0^2 \cos^2 \theta(x)     &  \frac{\sqrt{M_0^4 \, \cos^2 2\theta(x) - m^4_0}}{2} \\[2mm]
-\frac{\sqrt{M_0^4 \, \cos^2 2\theta - m^4_0}}{2}   &  M_0^2 \sin^2 \theta(x)
\end{array}
\right)\nonumber\\
\label{eq:m:vs:theta}
\eeqn
has the globally constant eigenvalues~\eq{eq:Mpm:phys}.

With constant $\theta$, the mass matrix~\eq{eq:m:vs:theta} corresponds to the same physical theory with the same spectrum. The matrices with different $\theta$, say $\theta_1$ and $\theta_2$, are related by the similarity transformation with $\xi = \theta_2 - \theta_1$. As we will see below, this statement is no longer valid for nonuniform $\theta \equiv \theta(x)$. We call the theories, related via a spacetime-dependent similarity transformation, self-similar theories. 

The unbroken $\PT$ symmetry requires that $M_0^2 \geqslant m_0^2$ and $\cos^2 2 \theta \geqslant m^4_0/M^4_0$,
implying that 
\beqn
- \theta_{\mathrm{max}} \leqslant  \theta \leqslant \theta_{\mathrm{max}}\,, 
\qquad
\theta_{\mathrm{max}} = \frac{1}{2} \arccos \frac{m_0^2}{M_0^2}\,.
\label{eq:theta:bound}
\eeqn
Then the parameter in the transformation matrix~\eq{eq:cS:xi} takes the simple form
\beqn
\kappa(x) = \frac{M_0^2}{m_0^2} \cos 2\theta(x) \qquad \mbox{with} \qquad  \kappa(x) \geqslant 1\,.
\label{eq:xi:theta}
\eeqn
Thus, the similarity transformation~\eq{eq:S2:global}, which diagonalizes the coordinate-dependent mass matrix~\eq{eq:m:vs:theta}
\beqn
\cS^{-1}\bigl(\xi_{\rm H}(\theta)\bigr) M^2(\theta) \cS \bigl(\xi_{\rm H}(\theta)\bigr) & = &
\begin{pmatrix}
M_+^2 & 0 \\
0 & M_-^2  \\
\end{pmatrix} \nonumber\\
& \equiv & \frac{1}{2} M_0^2 \bbbone + \frac{1}{2} m_0^2 \sigma_3\nonumber\\
\eeqn
is governed by the transformation parameter
\beqn
\xi_{\rm H}(x) = \frac{1}{2} \mathrm{arccosh}\,\kappa\bigl(\theta(x)\bigr)\,,
\label{eq:eta:H:theta}
\eeqn
with $\kappa$ given in Eq.~\eq{eq:xi:theta}.

\begin{figure}[!htb]
\begin{center}
\includegraphics[width=80mm, clip=true]{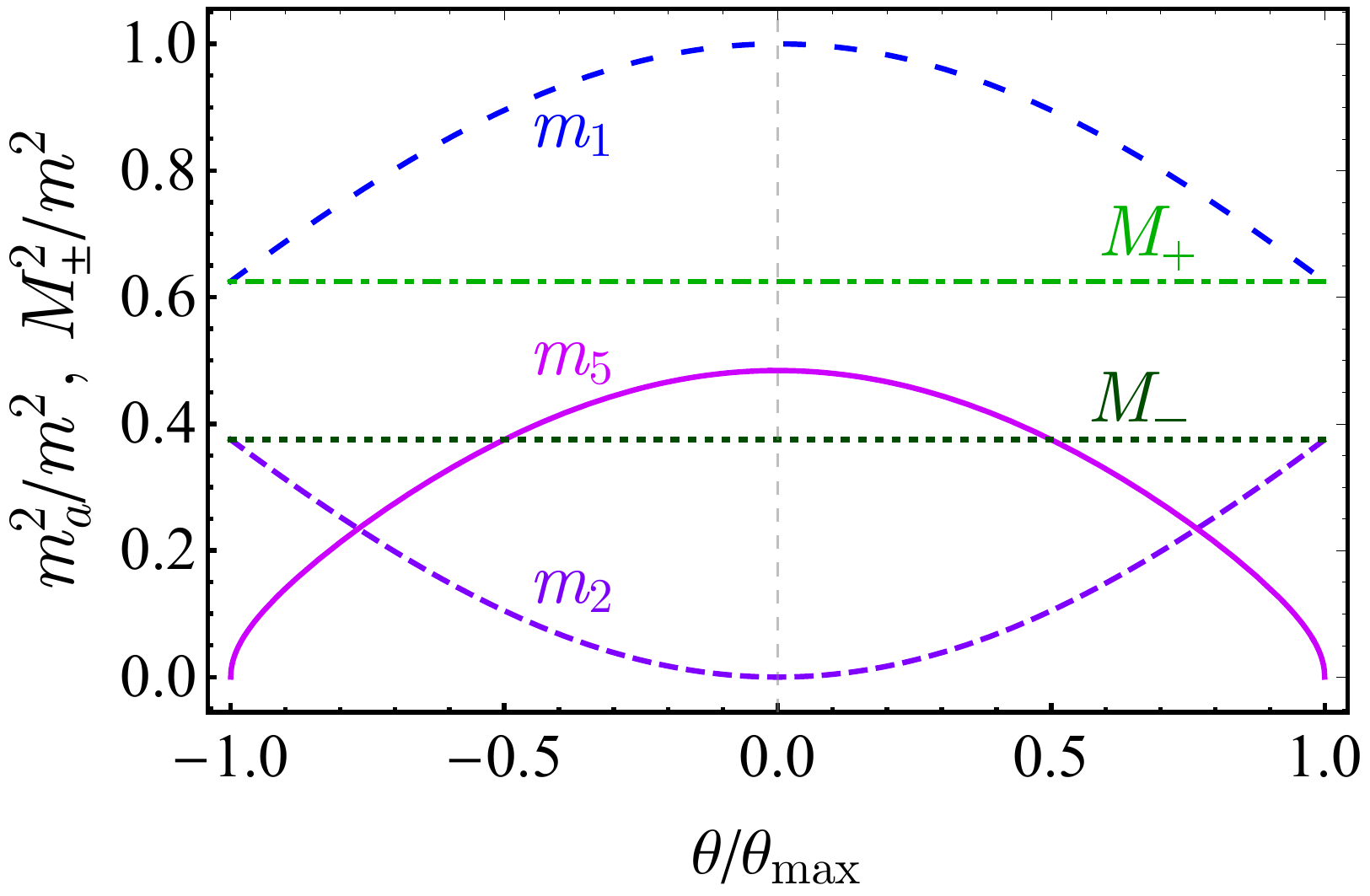} 
\end{center}
\caption{Dependence of the elements $m_1^2$, $m_2^2$, and $m_5^2$ of the squared mass matrix~\eq{eq:m:vs:theta} on the parameter $\theta$, which keeps the physical masses~\eq{eq:M:pm} $M_\pm$ constant in the non-Hermitian model~\eq{eq:L:prime}. The maximal angle $\theta_{\mathrm{max}}$ is defined in Eq.~\eq{eq:theta:bound}.}
\label{fig:masses}
\end{figure}

We can choose the parameter $\theta = \theta(x)$ in the form of an arbitrary function of the spacetime coordinate $x$, satisfying the bound~\eq{eq:theta:bound}. In Fig.~\ref{fig:masses}, we show the dependence of the entries in the squared mass matrix on the parameter $\theta$, which keeps the eigenvalues of the squared mass matrix~\eq{eq:M:pm} constant. While the squared mass matrix evolves in space or time---the latter dependence is encoded in the function $\theta = \theta(x)$---the physical spectrum of the theory remains untouched. 

As we show below, the spacetime inhomogeneity of the squared mass matrix can be encoded in the form of a vector gauge transformation in the isospace that acts on the upper and lower components of the doublet field $\Phi$. This mapping in the similarity space leads to the appearance of the similarity gauge field, discussed earlier. For the non-Hermitian theory, and while the eigenvalues of the squared mass matrix are constant, even a weak
local similarity field strongly affects the physical properties of the system, leading to instabilities for high-momentum modes, as we describe in the next section. In Sec.~\ref{sec:Hermitiancomparison}, we will discuss an {\it ab initio} Hermitian realization of the doublet scalar model, where the similarity field can also lead to an instability. However, in sharp contrast to the non-Hermitian model, the similarity field in the Hermitian model must be sufficiently strong in order to generate an instability, which makes the Hermitian case much less attractive from the point of view of phenomenology.

To obtain the would-be Hermitian counterpart of the non-Hermitian model after the local similarity transformation, it is enough to make the following substitutions in the generic non-Hermitian Lagrangian~\eq{eq:L:sim}:~$m_1 \to M_+$, $m_2 \to M_-$, and $m_5 \to 0$; and use the similarity gauge field 
\beqn
C^{\mu}_{\rm H} = - \partial^\mu \xi_{\rm H}\,,
\label{eq:C:NH}
\eeqn
where the similarity gauge parameter $\xi_{\rm H}$ is given by Eqs.~\eq{eq:xi:theta} and \eq{eq:eta:H:theta}. Alternatively, one can perform the local similarity transformation with the vanishing similarity gauge field $C_\mu$ in the original non-Hermitian basis; the results are the same. These procedures lead to the model
\beqn
\label{eq:L:sim:rotated} 
{\cal L}_{C,{\rm H}} & = & \partial_\mu \tilde{\phi}_+^* \partial^\mu \phi_+ - M_+^2\tilde{\phi}^*_+\phi_+ + \partial_\mu \tilde{\phi}_-^* \partial^\mu \phi_- - M_-^2\tilde{\phi}^*_- \phi_- \nonumber \\
& & + C^\mu_{\rm H} C_{{\rm H},\mu} (\tilde{\phi}^*_+ \phi_+ + \tilde{\phi}^*_- \phi_- ) +  C_{{\rm H},\mu} J^\mu_{\rm H}\,,
\eeqn
where $J^{\mu}_{\rm H}$ is the non-Hermitian similarity current given by Eq.~\eq{eq:J:sim} with $C_\mu = C_{{\rm H},\mu}$, \smash{$\phi_{1(2)}\to\phi_{+(-)}$} and \smash{$\tilde{\phi}^*_{1(2)}\to \tilde{\phi}^*_{+(-)}$}. The fields $\phi_\pm$ correspond to the diagonal mass entries $M_\pm$. 

The fact that the current $J^{\mu}_\NH$ remains non-Hermitian means that the local similarity transformation does not map the Lagrangian to an Hermitian one. It is for this reason that the tildes persist on the conjugated fields~$\tilde{\phi}_{\pm}^*$.

The subscript ``H'' in the similarity gauge field~$C_\NH^\mu$, given in Eq.~\eq{eq:C:NH}, is used to stress that this similarity field is a pure gauge field, which arises from a diagonalization of the coordinate-dependent squared mass matrix of the two-scalar-field model~\eq{eq:L:boson} with a vanishing vector field $C_\mu=0$. In other words, the non-Hermitian two-scalar model with (i) a vanishing similarity field, $C_\mu = 0$, (ii) a coordinate-dependent squared mass matrix, but (iii) coordinate-independent eigenmasses can be transformed to a two-scalar model with (i${}^\prime$) coordinate-independent masses and (ii${}^{\prime}$) the pure-gauge similarity vector field~$C_\NH^\mu$. The similarity field enters the non-Hermitian Lagrangian~\eq{eq:L:sim} as $C^\mu \equiv C_\NH^\mu$ in the covariant derivatives \eq{eq:D:mu} and \eq{eq:D:tilde:mu}.

If the original mass matrix in the non-Hermitian Lagrangian is spacetime independent, the similarity gauge field vanishes in the would-be Hermitian representation~\eq{eq:L:sim:rotated}, i.e., $C_{\NH,\mu} = 0$, and the Hermitian model splits into two independent scalar theories with globally constant masses.  In this case, the equations of motion for $\tilde{\phi}_{\pm}^*$ reduce to the complex conjugates of the equations of motion for $\phi_{\pm}$, such that we can omit the tildes.


\subsection{Equations of motion and background fields}
\label{sec:eoms}

The classical equations of motion following from the Lagrangian~\eq{eq:L:sim:2} with a generic similarity gauge field $C^\mu$ read as follows:
\beqs
\beqn
& & \left( \Box + C^2 + m_1^2  \right) \phi_1  \nonumber \\
& & \hskip 18mm + \left[ (\partial \cdot C) + 2 (C \cdot \partial) + m_5^2  \right] \phi_2 = 0\,, \qquad \\
& & \left( \Box + C^2 + m_2^2  \right) \phi_2 \nonumber \\
& & \hskip 18mm + \left[(\partial \cdot C) + 2 (C \cdot \partial) - m_5^2  \right] \phi_1 = 0\,, \qquad \\
& & \left( \Box + C^2 + m_1^2  \right) {\tilde \phi}_1 \nonumber \\
& & \hskip 18mm - \left[(\partial \cdot C) + 2 (C \cdot \partial) + m_5^2  \right] {\tilde \phi}_2 = 0\,, \qquad \\
& & \left( \Box + C^2 + m_2^2  \right) {\tilde \phi}_2 \nonumber \\
& & \hskip 18mm - \left[(\partial \cdot C) + 2 (C \cdot \partial) - m_5^2  \right] {\tilde \phi}_1 = 0\,. \qquad 
\eeqn
\label{eq:EOS:2}
\eeqs
We use the dot ``$\cdot$'' to denote a scalar product both in four ($C^2 \equiv C \cdot C = C_\mu C^\mu$) and three (${\bs C}^2 \equiv {\bs C} \cdot {\bs C}$) dimensions. In the absence of the similarity field, $C_\mu = 0$, Eq.~\eq{eq:EOS:2} reduces to the equations of motion~\eq{eq:classical:free}.

Let us consider the effect of the similarity gauge field in Eq.~\eq{eq:EOS:2} on the spectrum of the non-Hermitian model, focusing first on the case of a constant (spacetime-independent) gauge field $C_\mu$. We can then use the plane-wave basis \smash{$\phi_a(x^\mu) = \phi_{a}(0) e^{- i k_\mu x^\mu}$, with $k^\mu = (\omega, {\bs k})$} and $a=1,2$, to determine the energy spectrum $\omega = \omega_{\bs k}$ as a function of the three-momentum~${\bs k}$. The modes of the tilded fields are obtained via the transformation~\eq{eq:PT:transformations:NH}. All four equations in Eq.~\eq{eq:EOS:2} give the same relation for the energy spectrum:
\beqn
& & (\omega^2 - {\bs k}^2 - C^2 - m_1^2) (\omega^2 - {\bs k}^2 - C^2 - m_2^2) \nonumber \\
& & \hskip 25mm + 4 \left( C^0 \omega - {\bs k} \cdot {\bs C} \right)^2 + m_5^4 = 0\,.
\label{eq:omega:det}
\eeqn
While it is not immediately obvious that there is no coordinate dependence to this expression, given the presence of the squared mass parameters $m_1^2$, $m_2^2$ and $m_5^2$, we will see that this is indeed the case in what follows.

Equation~\eqref{eq:omega:det} is a fourth-order algebraic equation, the solutions of which are rather cumbersome. However, making use of its Lorentz covariance, which originates from the relativistic nature of the plane waves, we can simplify the solutions of Eq.~\eq{eq:omega:det}. Depending on the timelike $(C^2 \equiv C_0^2 - {\bs C}^2 >0)$ or spacelike $(C^2 < 0)$  nature of the field $C_\mu$, we can use Lorentz boosts to bring the system to the frame in which the field $C^\mu$ is perfectly timelike, $C^\mu = (C^0, {\bs 0})$, or perfectly spacelike, $C^\mu = (0, {\bs C})$, respectively.

The energy spectrum of the timelike $C^\mu = (C_0, {\bs 0})$ field has the form
\beqn
& & \omega_{\pm,{\bs k}}^2 = {\bs k}^2 - C_0^2 + \frac{1}{2} \left( m_1^2 + m_2^2 \right) \nonumber
\label{eq:omega:C0}\\
& & \quad \pm \frac{1}{2} \sqrt{(m_1^2 - m_2^2)^2 - 4 m_5^4 - 8 C_0^2 (m_1^2 +m_2^2)- 16 C_0^2 \, {\bs k}^2} \nonumber \\
& & \equiv  {\bs k}^2 - C_0^2 + \frac{1}{2} \left( M_+^2 + M_-^2 \right)  \nonumber \\
& & \quad \pm \frac{1}{2} \sqrt{\left( M_+^2 - M_-^2 \right)^2  - 8 C_0^2 \left( M_+^2 + M_-^2 \right) - 16 C_0^2 \, {\bs k}^2}\,,\nonumber\\
\eeqn
which differs from the usual relativistic spectrum of the standard form $\omega_{\bs k}^2 = {\bs k}^2 + m^2$, provided $C^0 \equiv C_0 \neq 0$. Interestingly, we still have a non-Hermitian theory in the limit $m_5\to 0$, so long as $C^0\neq 0$. Note also that since \smash{$\omega_{\pm,{\bs k}}^2$} can be written entirely in terms of the eigenvalues of the squared mass matrix \smash{$M_{\pm}^2$}, both of which are coordinate independent, the frequencies themselves are also coordinate independent, as indicated earlier.

The energy spectrum~\eq{eq:omega:C0} demonstrates that the presence of the similarity gauge field leads to instabilities in the system. At zero momentum, i.e., ${\bs k} = {\bs 0}$, the instability does not occur provided the magnitude of the field $C_0$ satisfies the following three requirements:
\beqs
\beqn
m_1^2 + m_2^2 - 2 C_0^2 & \geqslant & 0\,, 
\label{eq:stability:C0:1}\\[1mm]
m_1^2 m_2^2 + m_5^4 + C_0^2 (m_1^2 + m_2^2) + C_0^4 & \geqslant & 0\,,
\label{eq:stability:C0:2}\\
\left( m_1^2 - m_2^2\right)^2 - 4 m_5^4 - 8 C_0^2 (m_1^2 + m_2^2) & \geqslant & 0\,.
\label{eq:stability:C0:3}
\eeqn
\label{eq:stability:NH:C0} 
\eeqs
In the absence of the field, i.e., $C_0 = 0$, these conditions reduce to those in Eq.~\eq{eq:stability:NH}. Assuming that the system is stable at $C_0 = 0$, we find that Eq.~\eq{eq:stability:C0:2} is satisfied automatically, while the two other requirements, Eqs.~\eq{eq:stability:C0:1} and \eq{eq:stability:C0:3}, can be combined into one simple relation
\beqn
C_0^2 \leqslant {\mathrm{min}} \left( \frac{m_1^2 + m_2^2}{2}, \, \frac{\left(m_1^2 - m_2^2\right)^2 - 4 m_5^4}{8(m_1^2 + m_2^2)}\right)\,.
\eeqn
Significantly, the instability always arises in the ultraviolet region. The system is stable provided the momentum ${\bs k}$ does not exceed a certain critical scale; specifically,
\beqn
{\bs k}^2 \leqslant k_c^2 = \frac{\left(m_1^2 - m_2^2\right)^2 - 4 m_5^4}{16 C_0^2} - \frac{m_1^2 + m_2^2}{2}\,.
\label{eq:stability:C0}
\eeqn

\begin{figure}[!htb]
\begin{center}
\includegraphics[width=80mm, clip=true]{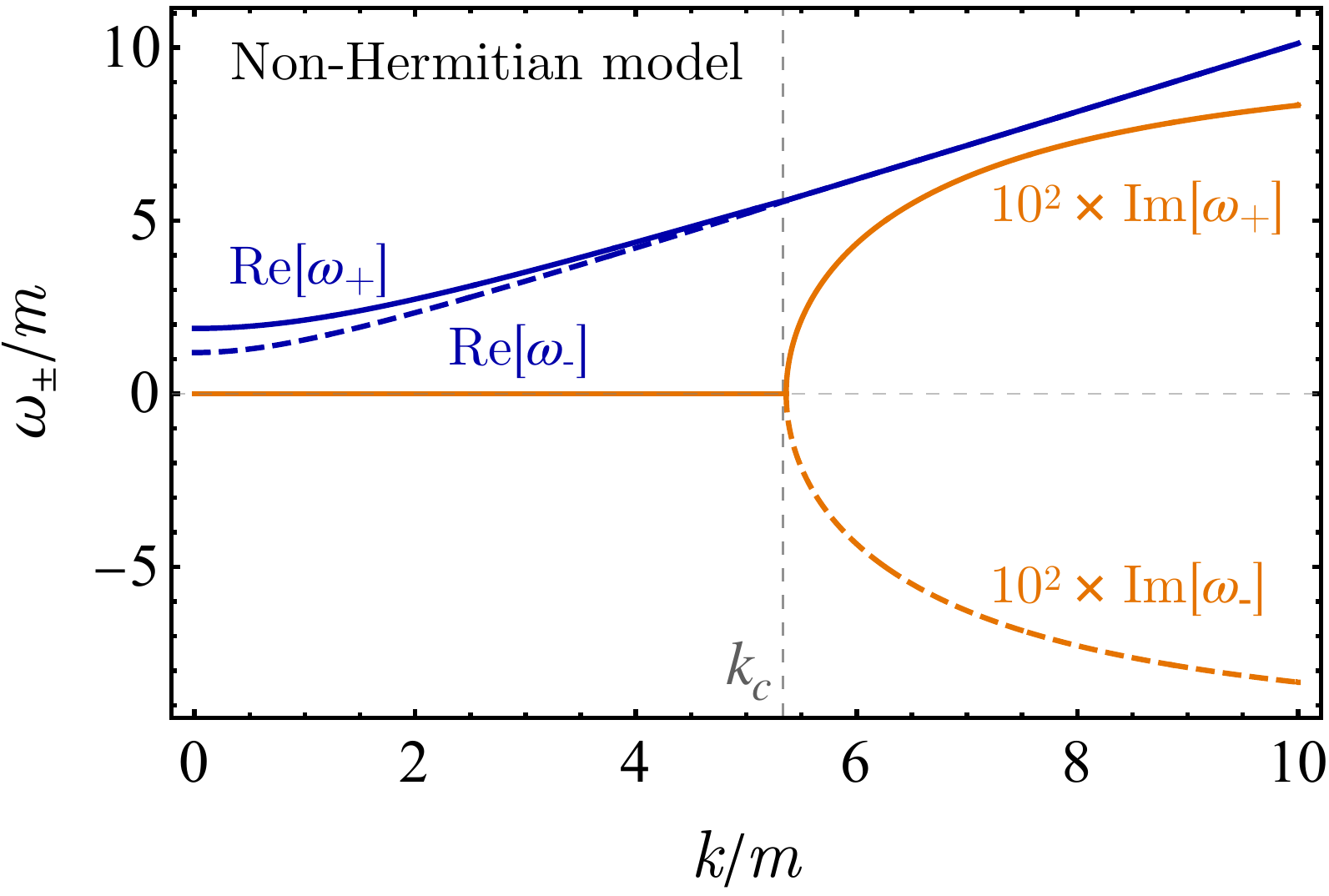} \\[1mm]
(a) \\[3mm]
\includegraphics[width=80mm, clip=true]{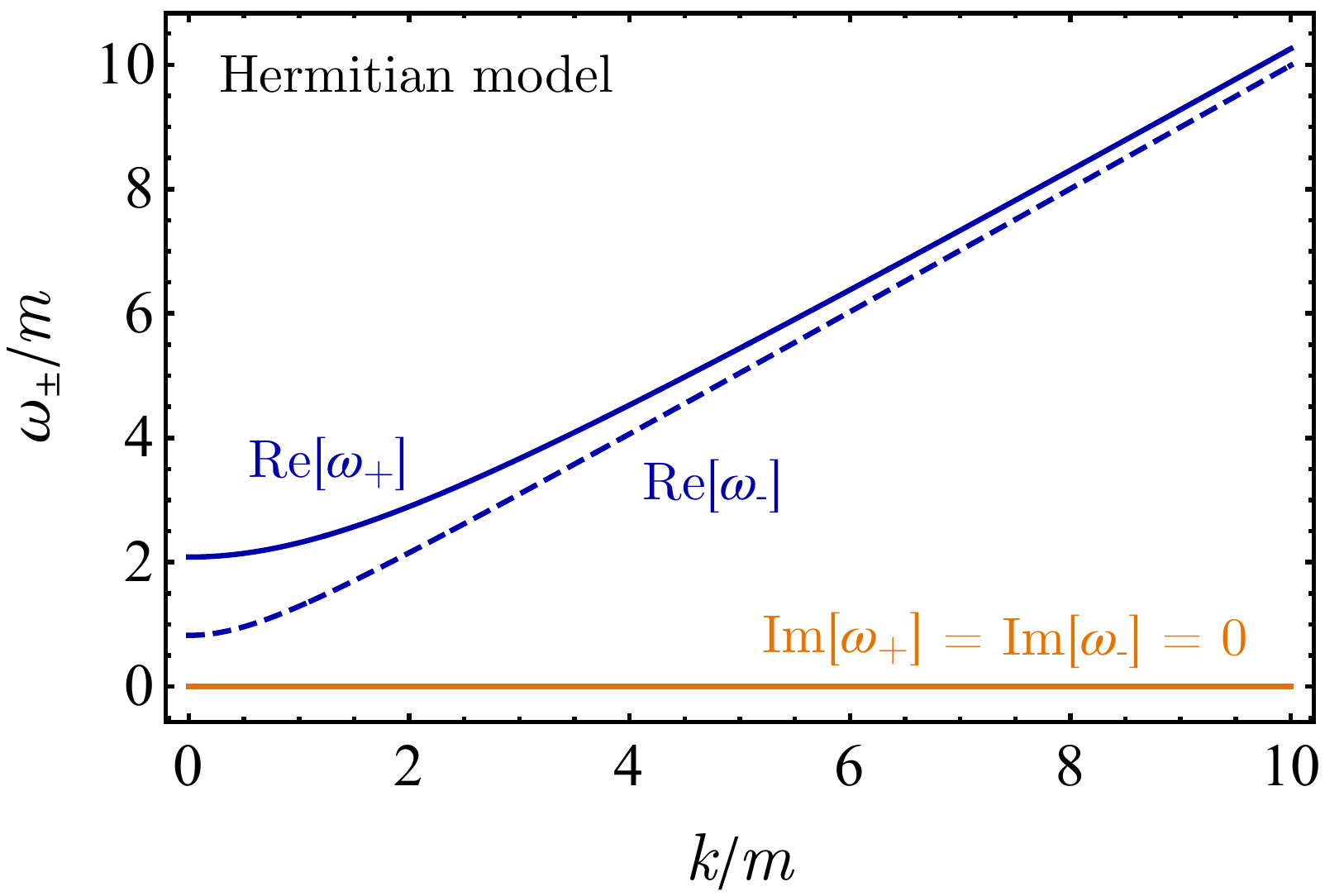} \\[1mm]
(b)\\[-3mm]
\end{center}
\caption{(a) The real and imaginary parts of the energy spectrum~\eq{eq:omega:C0} as a function of momentum $k = |{\bs k}|$ for the timelike similarity gauge field $C^\mu = (C^0,{\bs 0})$ and the parameters $m_1 = m_2/2 = m_5 = m$ and $C^0 = m/10$. The imaginary part is multiplied by a factor of $10^2$ to increase its visibility. The similarity gauge field makes the high-energy modes unstable at momenta larger than the critical value $k_c$, Eq.~\eq{eq:stability:C0}, without affecting the stability of the lower-energy modes, including the ground state. (In this example, $k_c \simeq 5.36 m$.) Notice that the real parts of the frequencies $\omega_{\pm}$ become degenerate for momenta $k>k_c$. (b) The corresponding plots for the {\it ab initio} Hermitian model of Sec.~\ref{sec:Hermitiancomparison} with the identical parameters. The energy spectrum is 
real for any value of momentum $k$.}
\label{fig:omega:C0}
\end{figure}

The energy dispersion for the timelike similarity field is illustrated in Fig.~\ref{fig:omega:C0}(a). The emergence of the instability is clearly seen at large values of momentum as determined by Eq.~\eq{eq:stability:C0}. For given mass parameters, the critical momentum scale $k_c$ determines the location of the exceptional points:~modes with momentum below this scale have real squared energies and lie in the regime of unbroken $\mathcal{PT}$ symmetry; modes with momentum above this scale have imaginary squared energies and lie in the regime of broken $\mathcal{PT}$ symmetry. We reiterate that Eqs.~\eq{eq:stability:NH:C0} to~\eq{eq:stability:C0} are all coordinate independent, since the only combinations of coordinate-{\it dependent} squared mass parameters appearing are the coordinate-{\it independent} ones $M_0^2=(m_1^2+m_2^2)/2$ and $m_0^4=(m_1^2-m_2^2)^2-4m_5^4$ [see Eq.~\eq{eq:m0M0}]. We have chosen to write these expressions in terms of the coordinate-dependent parameters in order to make connection with the original non-Hermitian squared mass matrix.


In the case of a spacelike similarity field, $C^\mu = (0,{\bs C})$, the spectrum becomes anisotropic:
\beqn
\omega_{\pm,{\bs k}}^2 & = & {\bs k}^2 - {\bs C}^2 + \frac{1}{2} \left( m_1^2 + m_2^2 \right) 
\label{eq:omega:vC}\nonumber\\
& & 
\pm \frac{1}{2} \sqrt{(m_1^2 - m_2^2)^2 - 4 m_5^4 - 16 ({\bs k} \cdot {\bs C})^2}\nonumber\\
& \equiv & {\bs k}^2 - {\bs C}^2 + \frac{1}{2} \left( M_+^2 + M_-^2 \right) 
\nonumber\\
& & 
\pm \frac{1}{2} \sqrt{(M_+^2 - M_-^2)^2 - 16 ({\bs k} \cdot {\bs C})^2}\,.
\eeqn
The stability conditions at zero momentum are:
\beqs
\beqn
m_1^2 + m_2^2 - 2 {\bs C}^2 & \geqslant & 0\,, 
\label{eq:stability:vC:1}\\
(m_1^2 - {\bs C}^2) (m_2^2 - {\bs C}^2) + m_5^4& \geqslant & 0\,,
\label{eq:stability:vC:2}\\
\left( m_1^2 - m_2^2\right)^2 - 4 m_5^4 & \geqslant & 0\,,
\label{eq:stability:vC:3}
\eeqn
\label{eq:stability:NH:vC} 
\eeqs
and the stability region at high momentum is limited to
\beqn
(m_1^2 - m_2^2)^2 - 4 m_5^4 \leqslant 16 ({\bs k} \cdot {\bs C})^2\,.
\label{eq:stability:vC}
\eeqn

\begin{figure*}[htb]
\begin{center}
\includegraphics[width=160mm, clip=true]{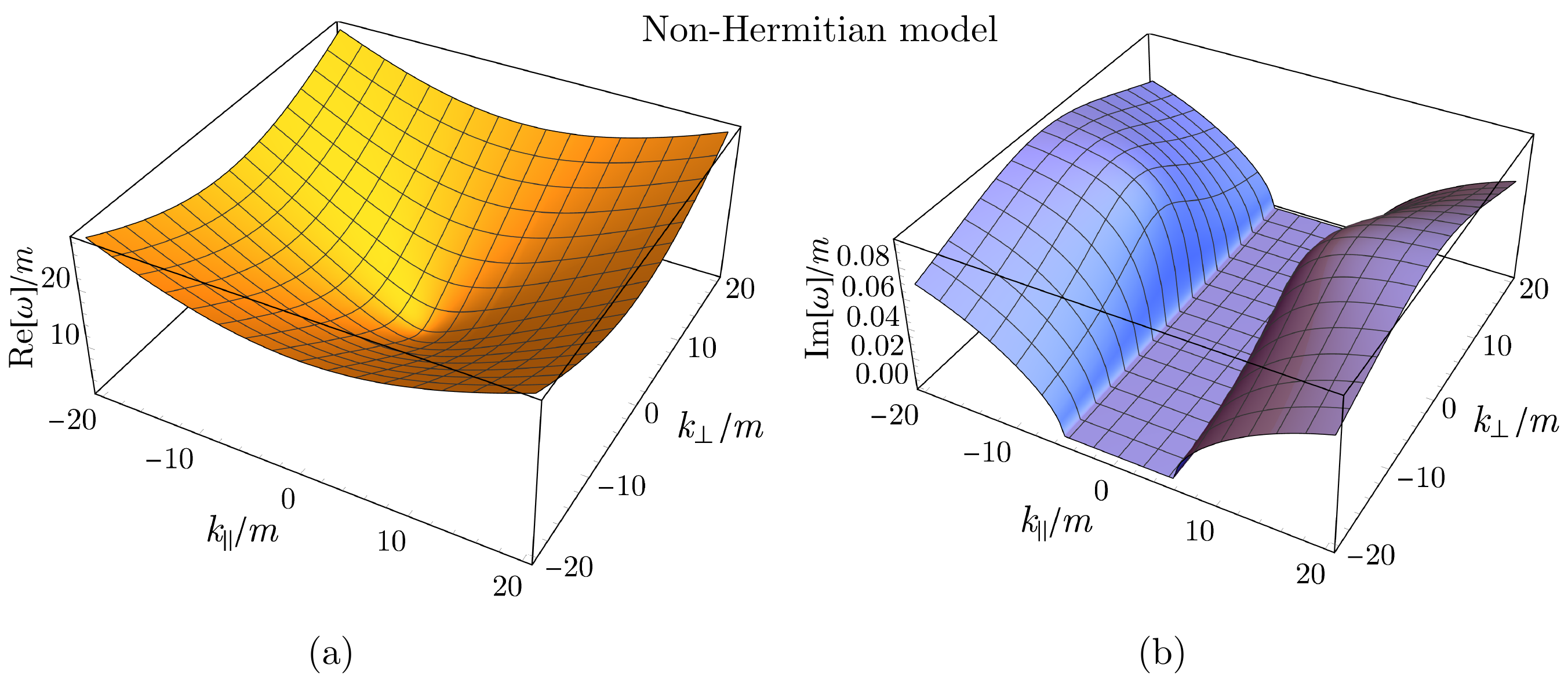} 
\end{center}
\vskip -5mm
\caption{The real (a) and imaginary (b) parts of the energy dispersion~\eq{eq:omega:vC} for the frequency $\omega = \omega_+$ with the model parameters $m_1 = m_2/2 = m_5 = m$ and $|{\bs C}| = m/10$ in the case of the spacelike similarity field $C^\mu = (0,{\bs C})$, with the momenta ${\bs k}_\|$ (${\bs k}_\perp$) being parallel (orthogonal) to the vector ${\bs C}$. The presence of the similarity gauge field leads to an instability of the high-energy modes propagating along the field direction. The stability of the lower-energy modes and the ground state is not affected. The frequency $\omega = \omega_-$ possesses a qualitatively similar real part, while the imaginary part comes with an opposite overall sign.}
\label{fig:omega:vC}
\end{figure*}

The real and imaginary parts of the energy dispersion~\eq{eq:omega:vC} for the spacelike similarity field $C^\mu = (0, {\bs C})$ are shown in Fig.~\ref{fig:omega:vC}. The stable region is determined by Eq.~\eq{eq:stability:vC}, which selects a strip in the longitudinal direction with respect to the field axis ${\bs k}_\| \| {\bs C}$. The transverse momenta are denoted ${\bs k}_\perp \perp {\bs C}$.

The presence of the instability for high-momentum modes might, at first sight, seem to cast doubts on the phenomenological viability of the model described in this work.  However, this model is understood to be an effective description, wherein the spacetime dependence of the squared mass parameters arises from interactions with other dynamical degrees of freedom that are not treated explicitly here. The instability for high-momentum modes indicates that this effective description breaks down, and it is then necessary to consider the dynamics of these additional degrees of freedom and the mechanism by which the spacetime dependence emerges.


\subsection{Comparison with the Hermitian model}
\label{sec:Hermitiancomparison}

It is helpful to compare the non-Hermitian model involving the similarity gauge field to an analogous construction for an {\it ab initio} Hermitian model composed of two complex scalar fields with a Hermitian mass mixing, i.e.,
\beqn
\mathcal{L}'&=&\partial_{\mu}\phi^*_1\partial^{\mu}\phi_1+\partial_{\mu}\phi_2^*\partial^{\mu}\phi_2\nonumber\\&&-m_1^2|\phi_1|^2-m_2^2|\phi_2|^2-m_5^2(\phi_1^*\phi_2+\phi_2^*\phi_1)\,.
\label{eq:L:Hermitian}
\eeqn
The squared Hermitian mass matrix 
\beqn
M^{\prime 2} = 
\begin{pmatrix}
m_1^2 & m_5^2 \\[1mm]
m_5^2 & m_2^2
\end{pmatrix}
\label{eq:mass:H}
\eeqn
is diagonalized by an $SO(2)$ transformation of the form
\beqn
    M^{\prime2}\to U^{\dag}M^{\prime2}U\,,
\eeqn
with
\beqn
U = e^{-i\sigma_2\xi'},\qquad \xi'=\frac{1}{2}\arctan\frac{2m_5^2}{m_1^2-m_2^2}\,,
\label{eq:U:Hermitian}
\eeqn
where $\sigma_2$ is the second Pauli matrix.
Notice that this is nothing other than the analytic continuation $m_5^2\to \pm i m_5^2$ of the non-Hermitian model.

However, if we take $m_{1,2,5}^2=m_{1,2,5}^2(x)$ and demand that the eigenmasses
\beqn
M_{\pm}^{\prime2} & = & \frac{1}{2}\Bigl\{m_1^2(x)+m_2^2(x) \nonumber \\
& & \pm\sqrt{\bigl[m_1^2(x) - m_2^2(x) \bigr]^2 + 4 m_5^4(x)}\Bigr\}\,,
\label{eq:mass:phys:Hermitian}
\eeqn
are spacetime-independent quantities, the same arguments as for the non-Hermitian model lead us to the Lagrangian\\
\beqn
{\cal L}' &=& (D_\mu \Phi)^{\dag} D^\mu \Phi - \Phi^{\dag}M^{\prime 2}_{\rm diag}\Phi\nonumber\\
& = & \left(\partial_{\mu}\phi_+^*+C_{\mu}\phi_-^*\right)\left(\partial^{\mu}\phi_++C^{\mu}\phi_-\right)\nonumber\\&&+\left(\partial_{\mu}\phi_-^*-C_{\mu}\phi_+^*\right)\left(\partial^{\mu}\phi_--C^{\mu}\phi_+\right)\nonumber\\&&- M_+^{\prime 2}\phi^*_+\phi_+- M_-^{\prime 2}\phi^*_- \phi_-\,,
\label{eq:Hermitian:diagonalized} 
\eeqn
where
\beqn
D^{\mu}=\partial^{\mu}\mathbb{I}_2+i\sigma_2C^{\mu}
\eeqn
is the covariant derivative equipped with the Hermitian similarity gauge field 
\beqn
C_{\mu} = - \partial_\mu \xi'\,,
\label{eq:C:Hermitian}
\eeqn
which depends on the parameter $\xi'$ given in Eq.~\eq{eq:U:Hermitian}.

\begin{figure*}[!htb]
\begin{center}
\includegraphics[width=180mm, clip=true]{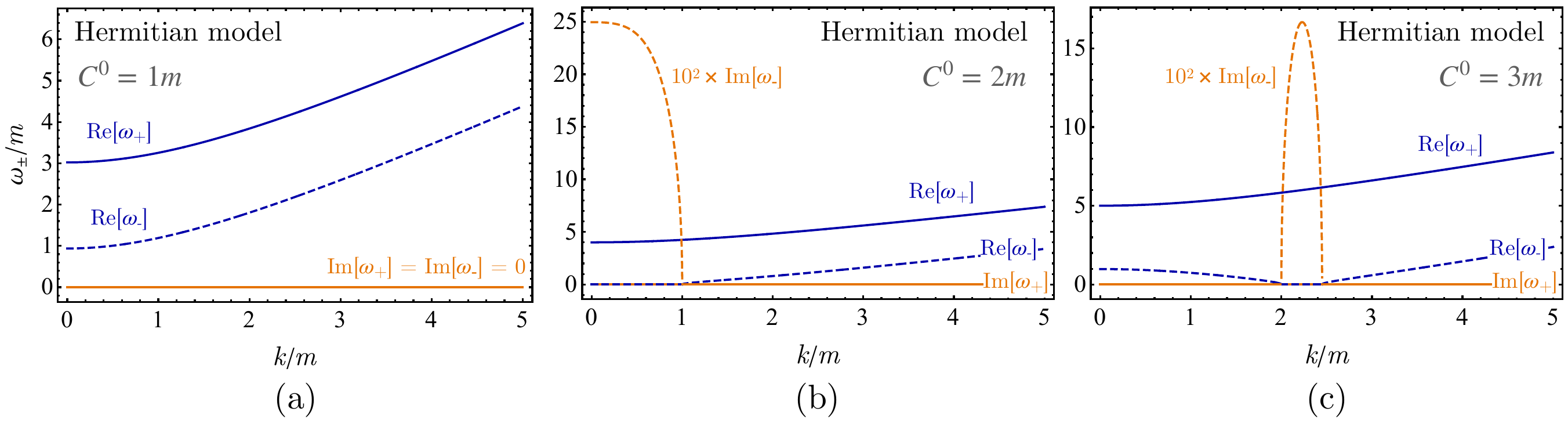} 
\end{center}
\vskip -5mm
\caption{The dispersion relations for the {\it ab initio} Hermitian model~\eq{eq:Hermitian:diagonalized} with the masses $m_1 = m_2 = 2m$ and $m_5 = m$ in the presence of the timelike similarity field $C^0 \equiv C_0$. The eigenvalues of the mass matrix, $M^\prime_+ = \sqrt{3} \simeq 1.7 m$ and $M^\prime_- = \sqrt{5} \simeq 2.2 m$, determine the three possible regimes of the model: (a) $C^0 = m < M^\prime_-$, a completely stable case; (b) $M^\prime_- < C^0 = 2 m < M^\prime_+$, the unstable region of the $\omega_-$ mode appears in the interior of the momentum surface ${\bs k}^2 < C_0^2 - M^{\prime 2}_-$; (c) $C^0 = 3m > M^\prime_+$, the unstable region of the same mode corresponds 
to a shell in the momentum space,
$C_0^2 - M^{\prime 2}_+ < {\bs k}^2 < C_0^2 - M^{\prime 2}_-$. The $\omega_+$ mode is always stable.}
\label{fig:omega:C0:H}
\end{figure*}

The conserved current is
\beqn
    J^{\mu}&=&\phi_-^*\partial^{\mu}\phi_+-\phi_+^*\partial^{\mu}\phi_-+(\partial^{\mu}\phi_+^*)\phi_--(\partial^{\mu}\phi_-^*)\phi_+\nonumber\\&&+2C^{\mu}(\phi_+^*\phi_++\phi_-^*\phi_-)\,,
\eeqn
and the equations of motion are as follows:
\beqs
\beqn
\left[\Box-C^2+M_+^{\prime 2}\right]\phi_++\left[\left(\partial\cdot C\right)+2\left(C\cdot \partial\right)\right]\phi_-&=&0\,,\nonumber\\&&\\
\left[\Box-C^2+M_-^{\prime 2}\right]\phi_--\left[\left(\partial\cdot C\right)+2\left(C\cdot \partial\right)\right]\phi_+&=&0\,.\nonumber\\&&
\eeqn
\eeqs
For a constant similarity field $C_{\mu}$, the energy spectrum is obtained from the equation
\beqn
&&\left(\omega^2-{\bs k}^2+C^2-M_+^{\prime2}\right)\left(\omega^2-{\bs k}^2+C^2-M_-^{\prime2}\right)\nonumber\\&&\qquad-4\left(C^0\omega-{\bs k}\cdot {\bs C}\right)^2=0\,.
\eeqn
Taking, as before, the purely timelike case $C^{\mu}=(C^0,{\bs 0})$, we get the following dispersion relation:
\beqn
& & \omega_{\pm,{\bs k}}^{\prime 2}={\bs k}^2+C_0^2+\frac{1}{2}\left(M_+^{\prime 2}+M_-^{\prime 2}\right)\nonumber\\& &\pm\frac{1}{2}\sqrt{\left(M_+^{\prime 2}-M_-^{\prime 2}\right)^2+8C_0^2\left(M_+^{\prime 2}+M_-^{\prime 2}\right)+16C_0^2{\bs k}^2}\,,\nonumber\\&&\equiv{\bs k}^2+C_0^2+\frac{1}{2}\left(m_1^{2}+m_2^{2}\right)\nonumber\\& &\pm\frac{1}{2}\sqrt{\left(m_1^{2}-m_2^{2}\right)^2+4m_5^4+8C_0^2\left(m_1^{2}+m_2^{2}\right)+16C_0^2{\bs k}^2}\,.\nonumber\\
\label{eq:C0:H}&&
\eeqn
The purely spacelike case $C^{\mu}=(0,{\bs C})$ gives us
\beqn
\omega_{\pm,{\bs k}}^{\prime 2} & = & {\bs k}^2+{\bs C}^2+\frac{1}{2}\left(M_+^{\prime 2}+M_-^{\prime 2}\right)\nonumber\\
& & \pm\frac{1}{2}\sqrt{\left(M_+^{\prime 2}-M_-^{\prime 2}\right)^2+16\left({\bs k}\cdot {\bs C} \right)^2}\,.\nonumber\\ 
& \equiv & {\bs k}^2+{\bs C}^2+\frac{1}{2}\left(m_1^{2}+m_2^{2}\right)\nonumber\\
& & \pm\frac{1}{2}\sqrt{\left(m_1^{2}-m_2^2 \right)^2+4m_5^4+16\left({\bs k}\cdot {\bs C}\right)^2}\,.\nonumber\\
\label{eq:vC:H}&&
\eeqn
Notice that the relations~\eq{eq:C0:H} and~\eq{eq:vC:H}, respectively, are again the analytic continuations of the non-Hermitian results in Eqs.~\eqref{eq:omega:C0} and~\eqref{eq:omega:vC}, with $m_5^2\to \pm im_5^2$ and $C^{\mu}\to \pm iC ^{\mu}$. This analytic continuation leads to a substantial difference between the Hermitian and non-Hermitian models.

Consider first the timelike case~\eq{eq:C0:H}. For a weak similarity field $C^0$, the energy spectrum of the Hermitian model is purely real, indicating the absence of any instability. This property is illustrated in Fig.~\ref{fig:omega:C0}(b), where the Hermitian parameters were taken to match the corresponding plots for the non-Hermitian model depicted in Fig.~\ref{fig:omega:C0}(a).

As the similarity field strengthens in the Hermitian model, modes in a window of wavelengths develop an instability. While the $\omega_+$ branch of the spectrum is always real, the instability emerges for the $\omega_-$ modes when the similarity field $C^0 \equiv C_0$ exceeds a critical value, given by the lowest physical  mass~\eq{eq:mass:phys:Hermitian}, and $\omega_-^2$ becomes negative. This occurs when
\beqn
|C^0| > C_c^0 = M^\prime_-\,.
\eeqn
The unstable modes, having a nonzero imaginary component in the energy dispersion, appear for the momenta
\beqn
\max \left(0, C^2_0 - M^{\prime 2}_+ \right) < {\bs k}^2 < C^2_0 - M^{\prime 2}_- \,,
\eeqn
while other modes are stable.

In the range of strengths $M^\prime_+ > |C^0| > M^\prime_-$, the instability occurs within the sphere $0 \leqslant {\bs k}^2 < C^2_0 - M^{\prime 2}_-$ in momentum space. If the similarity field exceeds the higher physical mass, i.e., $|C^0| > M^\prime_+$, then the instability takes place within the momentum shell $C^2_0 - M^{\prime 2}_+ < {\bs k}^2 < C^2_0 - M^{\prime 2}_-$. Notice that in the unstable region, the $\omega_-$ branches of the spectrum are zero modes in a sense that the real part of the energy is vanishing and the energy dispersion $\omega_-$ is a purely imaginary function of momentum.

The energy dispersion in the Hermitian model in the presence of the timelike similarity field $C^0$ is illustrated in Fig.~\ref{fig:omega:C0:H}.  Contrary to the plane-wave instability in the non-Hermitian model, the instability in the Hermitian model appears only for large values of the timelike field, as discussed in the caption of this figure.

The spacelike similarity field ${\bs C}$ leads to the dispersion relation~\eq{eq:vC:H}, which becomes complex if and only if the determinant of the Hermitian mass matrix~\eq{eq:mass:H} is negative. Therefore, the dispersion relation~\eq{eq:vC:H} develops a complex part provided one of the mass eigenvalues~\eq{eq:mass:phys:Hermitian} is purely imaginary even in the absence of the similarity field ${\bs C}$. This case corresponds to a trivial tachyonic instability of the {\it {ab initio}} Hermitian model~\eq{eq:L:Hermitian}, which is not interesting from the phenomenological point of view\footnote{The tachyonic instability can, however, be generated by the Englert-Brout-Higgs mechanism in an interacting model, which is not considered in our paper.}. One can also show that a nonvanishing spatial similarity field, contrary to its temporal analogue, improves the stability properties of the model by increasing the real-valued part of the energy squared $\omega_\pm^2$. This, the spatial similarity field, does not lead to any instability in the Hermitian model contrary to its non-Hermitian analogue. The latter requires the presence of only the tiniest similarity field $C_\mu$ to induce, according to Eqs.~\eq{eq:stability:C0} and~\eq{eq:stability:vC}, the instability of modes with sufficiently high energy while keeping moderate and low-energy states stable. This property makes the concept of the similarity gauge field in the non-Hermitian theory attractive from a phenomenological point of view in clear distinction with the Hermitian case.

Summarizing, we have seen that, while both Hermitian and non-Hermitian models possess an instability in the background of similarity gauge fields, there is a number of essential differences between the properties of these instabilities.

First, in the Hermitian case, the instability is realized at very strong background fields of the order of the mass of the particles, while the instability in the non-Hermitian model occurs at any value of the field.

Second, in the Hermitian model, the instability occurs within a finite window of momenta, typically of the order of the inverse Compton wavelength of the scalar particles. On the contrary, the unstable modes in the non-Hermitian model appear at very high energies with wavelengths much shorter that the Compton wavelength of the particle. We therefore observe that the non-Hermitian model features a novel IR/UV mixing, with a weak similarity field (corresponding to small, i.e., IR gradients of the mass parameters) leading to instabilities of the high-energy (UV) particle modes. The Hermitian model does not possess this IR/UV mixing.

These properties make the instability in the Hermitian model less useful from the point of view of present-day phenomenology, contrary to its non-Hermitian counterpart. Even so, the instability in the Hermitian case could still be important in the early Universe, where strong variations of the mass matrix could occur due to the presence of thin domain walls.


\section{Physical realization}
\label{sec:physical_realization}

In the non-Hermitian model, the symmetric $2 \times 2$ mass matrix involves three parameters~\eq{eq:mass:NH} that encode two physical masses~\eq{eq:M:pm}. Fixing the eigenvalues of the squared mass matrix, we still have one unfixed degree of freedom with which we can make the mass matrix spacetime dependent while keeping the eigenvalues globally constant in the whole spacetime. This behavior is reproduced in Fig.~\ref{fig:masses}, where the role of the auxiliary parameter is played by the angle $\theta$, which enters the squared mass matrix via Eq.~\eq{eq:m:vs:theta}.

In the case where the angle $\theta$ is a uniform and time-independent parameter, the choice of its value 
has no effect on the physical spectrum of the model. If the parameter $\theta$ is inhomogeneous, 
it leads to the appearance of a nonzero similarity field~\eq{eq:C:NH}, which affects the particle spectrum by modifying the dispersion relation and generating an instability at high energies. 

Let us consider the case where the entries of the mass matrix of the model in two distant spacetime regions are connected by a slowly varying similarity transformation, $\xi_{\NH} = - C_\mu x^\mu + O(x^2)$. Assuming that the variation is small, i.e., $C_0^2 \ll m^2$ and ${\bs C}^2 \ll m^2$, where $m$ defines the scale of the physical masses in the model, we expand the energies in powers of the similarity field and momenta to check the effect on the low-energy spectrum.

If the similarity field evolves in time but not in space, the rotation induces the temporal field $C_0 = - \partial_t \xi_\NH(t)$, producing the following correction to the energy~\eq{eq:omega:C0} of the long-wavelength modes:
\beqn
\omega_\pm^2 & = & M_\pm^2 - C_0^2 \left( 1 \pm 2 \frac{M_+^2 + M_-^2}{M_+^2 - M_-^2} \right) \nonumber \\
& & + {\bs k}^2 \left( 1 \mp \frac{4 C_0^2}{M_+^2 - M_-^2} \right) + O\left(C_0^4, {\bs k}^4\right)\,.
\label{eq:omega:C0:low}
\eeqn
The critical momentum, which determines the onset of the instability of the high-energy modes, is determined by Eq.~\eq{eq:stability:C0}:
\beqn
k_c = \frac{M_+^2 - M_-^2}{4 |C_0|} + O\left(C_0\right)\,.
\label{eq:kc:C0}
\eeqn
Notice that we always arrange the modes as $M_+ > M_0$.

For the spacelike inhomogeneity of the similarity gauge field, which induces the spatial field ${\bs C } = - {\bs \nabla} \xi_{\NH}$, we get the following low-energy expansion from Eq.~\eq{eq:omega:vC}:
\beqn
\omega_{\pm,{\bs k}}^2 & = & M_\pm^2 - {\bs C}^2 + {\bs k}_\perp^2
\label{eq:omega:vC:low}\nonumber \\
& & 
+ k_\|^2 \left( 1 \mp \frac{4 {\bs C}^2}{M_+^2 - M_-^2} \right) + O\left(C_0^4, {\bs k}^4\right)\,.
\eeqn
The critical momentum along the direction of the field ${\bs k}_\| \| {\bs C}$ comes from Eq.~\eq{eq:stability:vC}:
\beqn
k_{\|,c} = \frac{M_+^2 - M_-^2}{4 |{\bs C}|}\,.
\label{eq:kc:vC}
\eeqn
It has the same magnitude as the critical momentum~\eq{eq:kc:C0} for the timelike similarity field of the same value, i.e., $|{\bs C}| = |C_0|$. No instability appears at small values of the spatial field $\bs C$, as is illustrated in Fig.~\ref{fig:omega:C0}(b).

At the level of particle phenomenology, one can think about the field $\Phi$ as a generic doublet Higgs(-like) field. The effect of nonuniform similarity, which varies either in time or in space, has negligible consequences in the low-energy domain so that the inhomogeneous similarity can easily avoid detection. On the other hand, this phenomenon strongly affects the propagation of particles with very high energies.

For example, let us consider the inhomogeneous self-similar squared mass matrix, which varies with a similarity parameter of order unity, $\xi_\NH \sim O(1)$, at microscopic distances of 1 meter (or, equivalently, at the time scale of 1\,m/$c \simeq 3 \times 10^{-9}\,\mathrm{s}$, corresponding to the frequency of the order of 1 GHz). The similarity field has a minuscule magnitude $C = c\hbar/(1\,\mathrm{m}) \sim 2\times 10^{-7}\,\mathrm{eV}$ and its correction to the masses of particles, Eqs.~\eq{eq:omega:C0:low} and~\eq{eq:omega:vC:low}, lies well below the sensitivity of modern particle physics experiments at low energies. For particles with masses in the MeV range ($M_+ \sim M_- \sim M_+ - M_- > 0$), the particle instability appears at the critical momentum $k_c$, Eqs.~\eq{eq:kc:C0} and \eq{eq:kc:vC}, corresponding to energies $E_c = \hbar c k_c \simeq 10^{18} \, \mathrm{eV}$, which fall in the range of energies carried by ultra-high-energy cosmic rays. Of course, if the similarity effects vary more slowly (say, at the distance scale of 1 astronomical unit) then the low-energy mass corrections become even smaller while the high-energy cutoff, which marks the onset of the particle instability, increases. Therefore, the similarity evolution of the scalar field theory can rest unnoticed at low energy scales while substantially affecting the stability of scalar particles at high energy scales.


\section{Conclusions}
\label{sec:conc}

In the case of non-Hermitian field theories, the similarity transformation is usually understood as a global transformation acting in the space of fields that maps one field theory to another equivalent theory with precisely the same physical spectrum. Our article proposed to ``gauge'' the group of the similarity transformations, thus making the transformation dependent on the spacetime coordinate. In order to elucidate this point, we concentrated on both Hermitian and non-Hermitian field theories with two scalar fields.

The new similarity gauge symmetry leads to the emergence of a new type of vector field, which we called the similarity gauge field. The similarity gauge field acts as a gauge connection in the space of similar field theories characterized by the same (equivalent to a Hermitian) real-valued mass spectrum. 

The extension of the global similarity map to the local map leads to new effects for the particle properties. In our article, we considered the physically appealing case where the similarity gauge field is absent while the squared mass matrix of the two-field model is allowed to acquire coordinate dependence so that the local masses of particles are globally constant in the whole spacetime. This phenomenologically relevant setup leads to the appearance of a local similarity gauge field that, at the same time, keeps this ``locally similar'' model indistinguishable from a standard, low-energy scalar Hermitian model. 

In the {\it ab initio} Hermitian model, such coordinate dependence of the mass matrix leads to anisotropy in the propagation of particles and a tachyonic instability for a narrow window of momenta. On the other hand, in the non-Hermitian theory, we get several additional and principally new effects:

\begin{enumerate}

\item The high-energy particles become unstable at a particular wavelength determined by the strength of the similarity gauge field, which is related to the anisotropy of the mass matrix. These properties make our proposal phenomenologically interesting for ultra-high-energy particle physics, including detectable high-energy cosmic rays.

\item The emergent similarity gauge field keeps current low-energy phenomenology largely unaffected, thus making no experimentally detectable imprint on the low-energy spectrum over a wide range of reasonably chosen parameters.

\item The emergence of the similarity gauge field leads to a phenomenologically coherent interplay between the infrared and ultraviolet energy scales:~the lower the strength of the similarity gauge field, the more negligible the impact on the low-energy physics, including particle masses and anisotropy in particle propagation. At the same time, the weaker the similarity field, the higher the energy a particle should achieve to make the effects generated by the presence of the similarity field significant. (The latter effects include particle instabilities and anisotropies in particle propagation.) We stress that this behaviour arises only because anti-Hermitian interactions are permissible for a non-Hermitian theory. 

\item An elemental particle-physics model does not contain an inhomogeneous mass matrix as a fundamental quantity. Instead, the inhomegenity of the mass matrix should be considered as emerging from additional dynamics not considered here, e.g., as an effective operator that is parametrised in terms of the expectation value of some additional scalar field, $\chi$ say. The inhomogeneous mass matrix does not then correspond to the lowest-energy, vacuum state of the theory. Instead, when a particle with wavevector above the threshold set by the similarity gauge field propagates in the inhomogeneous background, it loses energy via emission of quanta of the field $\chi$. Since the total energy is conserved, one could expect that the radiation process excites the decaying eigenmode above threshold, slowing down the particle and, at the same time, smoothening the inhomogeneities in the expectation value of the $\chi$ field. The investigation of any such mechanism requires a separate study beyond the scope of the present work.

\item A distant analogue of the discussed phenomenon is the electromagnetic Cherenkov radiation that accompanies a highly-energetic particle entering a dielectric medium. The radiation occurs provided the magnitude of the particle wavevector exceeds a certain threshold, which is determined by the condition that the particle velocity equals the velocity of light in the medium. Eventually, the emitted radiation leads to a decrease in the particle energy, so that the wavevector reaches the critical value and the particle can no longer generate the radiation. This picture shares a similarity with the spectrum shown in Fig.~\ref{fig:omega:C0}(a) in the non-Hermitian model.

\end{enumerate}

An obvious extension of this article is to consider local similarity transformations of non-Hermitian fermionic field theories, such as those described in Refs.~\cite{Bender:2005hf, Jones-Smith:2009qeu, Alexandre:2015oha, Alexandre:2015kra, Chernodub:2017lmx, Alexandre:2020wki, Chernodub:2020cew}. We leave this for future work.


\acknowledgments

We thank Alberto Cortijo for initial collaboration on this paper. This work was supported by a Royal Society International Exchange [Grant No.~IES\textbackslash R3\textbackslash 203069]; a United Kingdom Research and Innovation (UKRI) Future Leaders Fellowship [Grant No.~MR/V021974/1]; and a Nottingham Research Fellowship from the University of Nottingham.


\appendix


\subsection*{Appendix:~Operator-level transformations}

Generalizing the transformations described in Ref.~\cite{Alexandre:2020gah}, the local similarity transformation $\mathcal{S}$ can be written in terms of the following operator:
\beqn
    &&\hat{S}(t,\mathbf{x})=\exp\Bigg[i \xi(t,\mathbf{x})\nonumber\\&&\qquad\times\int_{\mathbf{y}}\Big(\pi_1(t,\mathbf{y})\phi_2(t,\mathbf{y})+\pi_2(t,\mathbf{y})\phi_1(t,\mathbf{y})\nonumber\\&&\qquad-\tilde{\pi}^{\dag}_1(t,\mathbf{y})\tilde{\phi}^{\dag}_2(t,\mathbf{y})-\tilde{\pi}^{\dag}_2(t,\mathbf{y})\tilde{\phi}^{\dag}_1(t,\mathbf{y})\Big)\Bigg]\,.
\eeqn
We do not distinguish the operator-valued fields and conjugate momenta from their $c$-number counterparts so as to avoid further complicating our notation.

Making use of the canonical algebra~\cite{Alexandre:2020gah}
\beqs
\beqn
\Big[\phi_i(t,\mathbf{x}),\phi_j(t,\mathbf{y})\Big]&=&0\,,\\
\Big[\phi_i(t,\mathbf{x}),\pi_j(t,\mathbf{y})\Big]&=&i\delta_{ij}\delta^3(\mathbf{x}-\mathbf{y})\,,\\
\Big[\tilde{\phi}^{\dag}_i(t,\mathbf{x}),\tilde{\phi}^{\dag}_j(t,\mathbf{y})\Big]&=&0\,,\\
\Big[\tilde{\phi}^{\dag}_i(t,\mathbf{x}),\tilde{\pi}^{\dag}_j(t,\mathbf{y})\Big]&=&i\delta_{ij}\delta^3(\mathbf{x}-\mathbf{y})\,,\\
\Big[\phi_i(t,\mathbf{x}),\tilde{\phi}^{\dag}_j(t,\mathbf{y})\Big]&=&0\,,\\
\Big[\phi_i(t,\mathbf{x}),\tilde{\pi}^{\dag}_j(t,\mathbf{y})\Big]&=&0\,,
\eeqn
\eeqs
where $i,j=1,2$, we can show that the fields transform as
\beqs
\beqn
\phi_i(x)&\to& \phi_i(x)\cosh\xi(x)-\phi_{\slashed{i}}(x)\sinh\xi(x)\,,\\
\tilde{\phi}_i^{\dag}(x)&\to& \tilde{\phi}_i^{\dag}(x)\cosh\xi(x)+\tilde{\phi}_{\slashed{i}}^{\dag}(x)\sinh\xi(x)\,,
\eeqn
\eeqs
where $\slashed{i}=2$ if $i=1$ and vice versa. Hereafter, we omit the spacetime arguments for notational convenience. It then follows straightforwardly that
\beqs
\beqn
\partial_{\mu}\big[\hat{S}^{-1}\phi_i\hat{S}\big]&=& \cosh(\xi)\,\partial_{\mu}\phi_i-\sinh(\xi)\,\partial_{\mu}\phi_{\slashed{i}}\nonumber\\&&+\partial_{\mu}\xi\left[\sinh(\xi)\,\phi_i-\cosh(\xi)\,\phi_{\slashed{i}}\right]\,,\\
\partial_{\mu}\big[\hat{S}^{-1}\tilde{\phi}_i^{\dag}\hat{S}\big]&=&\cosh(\xi)\,\partial_{\mu}\tilde{\phi}^{\dag}_i+\sinh(\xi)\,\partial_{\mu}\tilde{\phi}^{\dag}_{\slashed{i}}\nonumber\\&&+\partial_{\mu}\xi\left[\sinh(\xi)\,\tilde{\phi}^{\dag}_i+\cosh(\xi)\,\tilde{\phi}^{\dag}_{\slashed{i}}\right]\,.
\eeqn
\eeqs

For $\xi=\text{const.}$, the kinetic terms are invariant under this transformation. Instead, for the local transformation, we have
\beqn
\label{eq:kinetic_term_transfo_1}
&&\sum_{i=1}^2\partial\tilde{\phi}_i^{\dag}\cdot\partial\phi_i\to\sum_{i=1}^{2}\partial\big[\hat{S}^{-1}\tilde{\phi}^{\dag}_i\hat{S}\big]\cdot \partial\big[\hat{S}^{-1}\phi_i\hat{S}\big]\nonumber\\&&\quad = \sum_{i=1}^2\left[\partial\tilde{\phi}_i^{\dag}\cdot \partial\phi_i-\tilde{\phi}_i^{\dag}\tilde{\phi}_i\left(\partial\xi\right)^2\right]\nonumber\\&&\quad +\partial\xi\cdot\left[\tilde{\phi}_1^{\dag}\partial\phi_2+\tilde{\phi}_2^{\dag}\partial\phi_1-\left(\partial\tilde{\phi}_1^{\dag}\right)\phi_2-\left(\partial\tilde{\phi}_2^{\dag}\right)\phi_1\right]\,,\nonumber\\
\eeqn
wherein we recognise the similarity current from Eq.~\eqref{eq:J:sim}.

Note that the transformation described here maps the Lagrangian but not the field operators themselves. In order to map both the Lagrangian and the field operators, it is necessary to construct the similarity transformation in Fock space and at the level of the particle and antiparticle creation and annihilation operators, as was done in Ref.~\cite{Alexandre:2020gah}. We refrain from doing so here, since the coordinate dependence of the squared mass matrix significantly complicates the Fock-space quantization for this model.


%

\end{document}